# Equation of state of boron subarsenide $B_{12}As_2$ to 47 GPa

Kirill A. Cherednichenko,[1] Yann Le Godec [2] and Vladimir L. Solozhenko [1,*]

[1] *LSPM–CNRS, Université Paris Nord, 93430 Villetaneuse, France*

[2] *IMPMC–CNRS, UPMC Sorbonne Universités, 75005 Paris, France*

## Abstract

*Compressibility of boron subarsenide $B_{12}As_2$ has been studied by synchrotron X-ray diffraction up to 47 GPa at room temperature in a diamond anvil cell using Ne pressure transmitting medium. A fit of experimental p-V data by Vinet equation of state yielded the bulk modulus of 150(4) GPa and its first pressure derivative of 6.4(3). No pressure-induced phase transitions have been observed.*

Keywords : *boron arsenide; high pressure; equation of state*

## I. Introduction

Boron-rich compounds attract considerable attention due to their unusual structure and properties. The $B_{12}$-icosahedral units are the common feature of these compounds. The majority of boron-rich solids have structures related to α-rhombohedral boron with corresponding interstitial atoms: $B_{12}C_3$, $B_{13}N_2$, $B_{12}O_2$, $B_{12}Si_{2-x}$, $B_{12}P_2$, etc. [1-5]. Owing to presence of electron-deficient three-center-two-electron bonds within icosahedra and covalent bonds linking adjacent icosahedra and interstitial atoms as the electron donors, the boron-rich compounds have outstanding mechanical and physical properties and can be used in many fields: from novel biological and chemical sensors to promising materials for armour [3,6,7].

Boron subarsenide $B_{12}As_2$ is a typical boron-rich compound, possessing a rather high bulk modulus [8-11] and interesting electronic properties (e.g. relatively high hole mobility, a wide energy band gap, etc.) [2,12-17]. It was found to be extraordinary sustainable to the radiation [17,18]: $B_{12}As_2$ showed no signs of damage even after prolonged electron bombardment and has an ability to recover from radiation damage. Such fortunate combination of properties makes boron subarsenide very promising semiconductor material for devices operated under extreme conditions, e.g. high temperature, high pressure, high radiation fields, etc.

\* Corresponding author: vladimir.solozhenko@univ-paris13.fr

Although 300-K phase stability of $B_{12}As_2$ has been studied up to 120 GPa by Raman spectroscopy [19,20], X-ray diffraction study of $B_{12}As_2$ has been performed to 25 GPa only [8], and methanol - ethanol mixture has been used as a pressure transmitting medium. From the fit of Birch-Murnaghan equation of state (EOS) to the experimental $p$-$V$ data the bulk modulus ($B_0$) of 216 GPa and its pressure derivative ($B_0'$) of 2.2 have been estimated. However, the theoretically predicted $B_0$-values of $B_{12}As_2$ are somewhat lower i.e. 181.7 GPa [9] and 197.6 GPa [10]. It should be noted that in the case of methanol - ethanol mixture the region of hydrostaticity is limited by pressure of about 10 GPa [21-24]. In other words, the reported data on $B_{12}As_2$ compressibility above 10 GPa are likely not precise. Thus, there are strong grounds to expect that $B_{12}As_2$ bulk modulus found in [8] is significantly overestimated, and hence more precise measurement of $B_{12}As_2$ compressibility is required. Here we report the equation-of-state data for boron subarsenide measured under quasi-hydrostatic conditions in the neon pressure transmitting medium up to 47 GPa at room temperature.

**II. Experimental**

Polycrystalline subarsenide $B_{12}As_2$ has been synthesized at 5.2 GPa and 2100 K by reaction of boron with molten arsenic in a toroid-type high-pressure apparatus with a specially designed high-temperature cell [25]. Amorphous boron (Johnson Matthey, 99%) and polycrystalline arsenic (Alfa Aesar, 99.9999%) powders were used as starting materials. Boron nitride (grade AX05, Saint-Gobain) capsules were used to isolate the reaction mixture from the graphite heater. X-ray powder diffraction analysis (G3000 TEXT Inel diffractometer, Cu $K\alpha 1$ radiation) showed that the recovered sample contains single-phase $B_{12}As_2$ with lattice parameters $a = 6.1353(2)$ Å and $c = 11.8940(7)$ Å that are in good agreement with literature data [8,26,27].

*In situ* X-ray diffraction experiments up to 47 GPa at room temperature were performed at Xpress beamline (Elettra) in a membrane diamond anvil cell with 300-μm culet anvils. The powder sample was loaded into a 100-μm hole drilled in a rhenium gasket pre-indented down to ~30 μm. Neon was used as a pressure transmitting medium to maintain quasi-hydrostatic conditions. The pressure was determined using equation of state of solid neon [28,29]; the maximum pressure uncertainty was estimated to be within ±0.6 GPa. The pressure drift at each pressure point did not exceed 0.3 GPa. High-brilliance synchrotron radiation from multipole superconducting wiggler was set to a wavelength of 0.4957 Å using a channel-cut Si (111) monochromator and focused down to 20 μm. MAR 345 image plate detector was employed to collect angle-dispersive X-ray diffraction patterns with exposure time of 600 seconds. The diffraction patterns were processed with FIT2D software [30]. The lattice parameters of $B_{12}As_2$ under pressure refined by Le Bail method (structureless whole profile fitting) using Maud [31] and PowderCell [32] software are presented in Table I.

## III. Results and discussion

The structure of $B_{12}As_2$ (Fig. 1) belongs to *R-3m* space group [27], similar to other boron-rich compounds with structures related to α-rhombohedral boron.

During compression the reflections of $B_{12}As_2$ monotonously shifted towards larger 2θ-values, and no evidence of any phase transition has been observed over the whole studied pressure range. Even at pressures above 40 GPa, the diffraction lines do not show observable broadening or weakening which is indicative of quasi-hydrostatic compression and absence of stresses in the sample (Fig. 2).

The one-dimensional analogue of the first-order Murnaghan equation of state [33] of the form

$$r = r_0 \left[1 + P\left(\frac{\beta'_{0,r}}{\beta_{0,r}}\right)\right]^{-\frac{1}{\beta'_{0,r}}} \quad (1)$$

where *r* is the lattice parameter (index 0 refers to ambient pressure), $\beta_{0,r}$ is the axial modulus and $\beta'_{0,r}$ is its pressure derivative, was used for approximation of the nonlinear relation between normalized lattice parameters and pressure (Fig. 3). The axis moduli that best fit the experimental data are $\beta_{0,a}$= 449±12 GPa and $\beta_{0,c}$= 461±17 GPa, and the corresponding pressure derivatives are $\beta'_{0,a}$= 21.7±1.1 and $\beta'_{0,c}$= 10.4±1.2. Thus, compression of $B_{12}As_2$ unit cell is slightly anisotropic with higher compressibility along *c*-axis.

Fig. 4 presents experimentally observed unit cell volume change under pressure. We employed Murnaghan [33] (Eq. 2), third-order Birch-Murnaghan [34] (Eq. 3) and Vinet [35] (Eq. 4) equations of state to fit the experimental data

$$P(V) = \frac{B_0}{B'_0}\left[\left(\frac{V}{V_0}\right)^{B'_0} - 1\right] \quad (2)$$

$$P(V) = \frac{3B_0}{2}\left[\left(\frac{V_0}{V}\right)^{\frac{7}{3}} - \left(\frac{V_0}{V}\right)^{\frac{5}{3}}\right]\left\{1 + \frac{3}{4}(B'_0 - 4)\left[\left(\frac{V_0}{V}\right)^{\frac{2}{3}} - 1\right]\right\} \quad (3)$$

$$P(V) = 3B_0\frac{(1-X)}{X^2}e^{(1.5(B'_0-1)(1-X))} \quad (4)$$

where $X = \sqrt[3]{\frac{V}{V_0}}$, $V_0$ is unit cell volume at ambient pressure.

The corresponding $B_0$ and $B_0'$ values are presented in Table II. All three equations of state well approximate the experimental *p-V* data, giving almost the same EOS parameters. However, the lower $\chi^2$ parameter for the Vinet EOS fit indicates a better applicability of this equation.

As it follows from Fig. 5, $B_{12}As_2$ has the lowest bulk modulus among all studied boron-rich pnictides. The first pressure derivative of bulk modulus (6.4) is very close to $B_0'$ values of $B_{12}P_2$ (5.5) [36] and $B_{12}O_2$ (6.0) [37]. Thus, the previously determined $B_0'$ value of 2.2 [8] and, hence, the corresponding $B_0$ value of 216 GPa seem to be incorrect. According to Wu et al. [8], $B_0 = 204$ GPa ($B_0'$ fixed to 4.0) of $B_{12}As_2$ is comparable with bulk moduli of boron allotropes α-$B_{12}$ [38] and β-$B_{106}$ [39] indicates that substitution of the $B_{10}$-B-$B_{10}$ chain with the arsenic atoms does not affect the compressibility. However, this statement contradicts the general tendency observed for boron-rich pnictides i.e. a larger covalent radius of interstitial atom ($r_{(N)} = 0.75$ Å, $r_{(P)} = 1.06$ Å, $r_{(As)} = 1.20$ Å) leads to a higher compressibility (Fig. 5). The results of the present work completely support this tendency. Summing up, the use of proper pressure transmitting medium allowed us to obtain the reliable values of bulk modulus and its pressure derivative for boron subarsenide and once again demonstrate the importance of maintaining quasi-hydrostatic conditions in EOS measurements.

**IV. Conclusions**

Compressibility of boron subarsenide, $B_{12}As_2$, has been studied at room temperature up to 47 GPa by synchrotron X-ray powder diffraction in a diamond anvil cell. The icosahedral *R*-3*m* structure of $B_{12}As_2$ is stable in the whole pressure range under study. The use of the proper pressure transmitting medium (Ne) allowed us to avoid non-hydrostaticity and, thus, to obtain reliable values of bulk modulus (150 GPa) and its first pressure derivative (6.4).

**Acknowledgements**

The authors thank Dr. V.A. Mukhanov (LSPM-CNRS) for the sample synthesis, Dr. A. Polian (IMPMC/SOLEIL) for help in DAC loading, and Dr. P. Lotti (Elettra) for assistance at Xpress beamline. Synchrotron X-ray diffraction experiments were carried out during beamtime allocated to Proposal 20160061 at Elettra Synchrotron (Trieste). This work was financially supported by the European Union's Horizon 2020 Research and Innovation Program under Flintstone2020 project (grant agreement No 689279).

Table 1. Experimental values of lattice parameters and unit cell volume of $B_{12}As_2$ (in hexagonal setting) as a function of pressure at room temperature.

| Pressure (GPa) | $a$ (Å) | $c$ (Å) | Volume (Å$^3$) |
|---|---|---|---|
| 0.0 | 6.1353(2) | 11.8940(7) | 387.72(5) |
| 6.0 | 6.0560(2) | 11.7299(6) | 372.55(4) |
| 12.1 | 6.0100(2) | 11.6380(7) | 364.04(5) |
| 17.7 | 5.9627(3) | 11.5150(5) | 354.54(5) |
| 21.1 | 5.9400(3) | 11.4640(7) | 350.29(6) |
| 22.4 | 5.9308(4) | 11.4420(6) | 348.54(7) |
| 27.8 | 5.9040(4) | 11.3470(7) | 342.53(7) |
| 30.1 | 5.8873(5) | 11.2905(7) | 338.89(8) |
| 35.2 | 5.8630(4) | 11.2553(8) | 335.05(7) |
| 37.6 | 5.8470(5) | 11.2126(8) | 331.96(8) |
| 40.5 | 5.8325(6) | 11.1670(8) | 328.98(9) |
| 43.5 | 5.8218(6) | 11.1500(9) | 327.27(9) |
| 47.2 | 5.8100(6) | 11.0900(11) | 324.19(10) |

Table 2. EOS parameters of $B_{12}As_2$ obtained from approximations of the experimental data using Vinet (*V*), Birch-Murnaghan (*B-M*) and Murnaghan (*M*) EOSs. $\chi^2$ is an indication of the fit quality (lower for a better fit).

| EOS | $B_0$ (GPa) | $B_0'$ | $\chi^2$ |
|---|---|---|---|
| *V* | 149.6±3.9 | 6.4±0.3 | 0.170 |
| *B-M* | 150.1±4.2 | 6.3±0.4 | 0.171 |
| *M* | 150.1±4.2 | 6.3±0.4 | 0.732 |

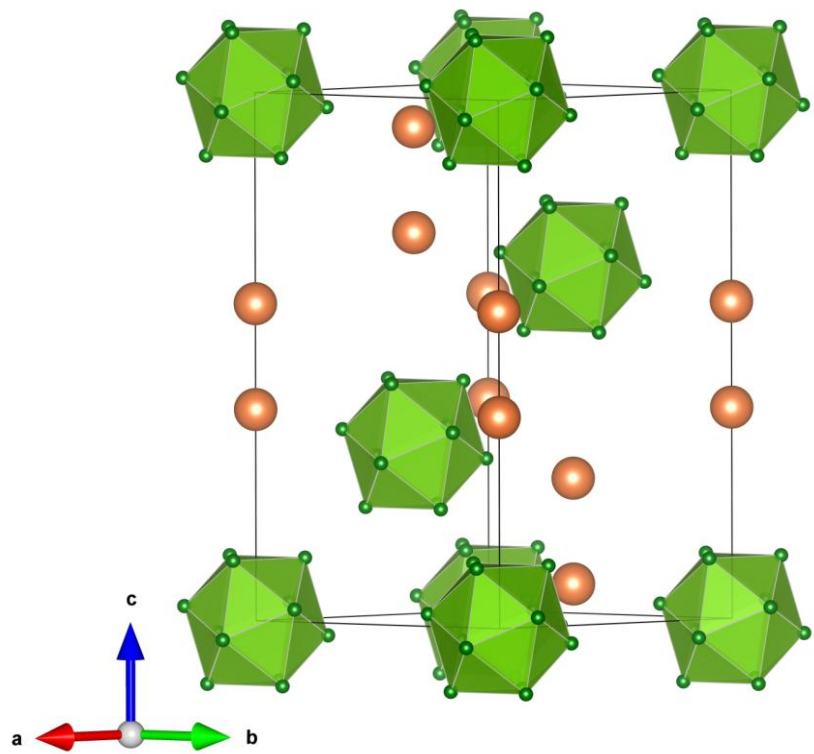

Fig. 1. Crystal structure of boron subarsenide $B_{12}As_2$ (green and orange balls are B and As atoms, respectively).

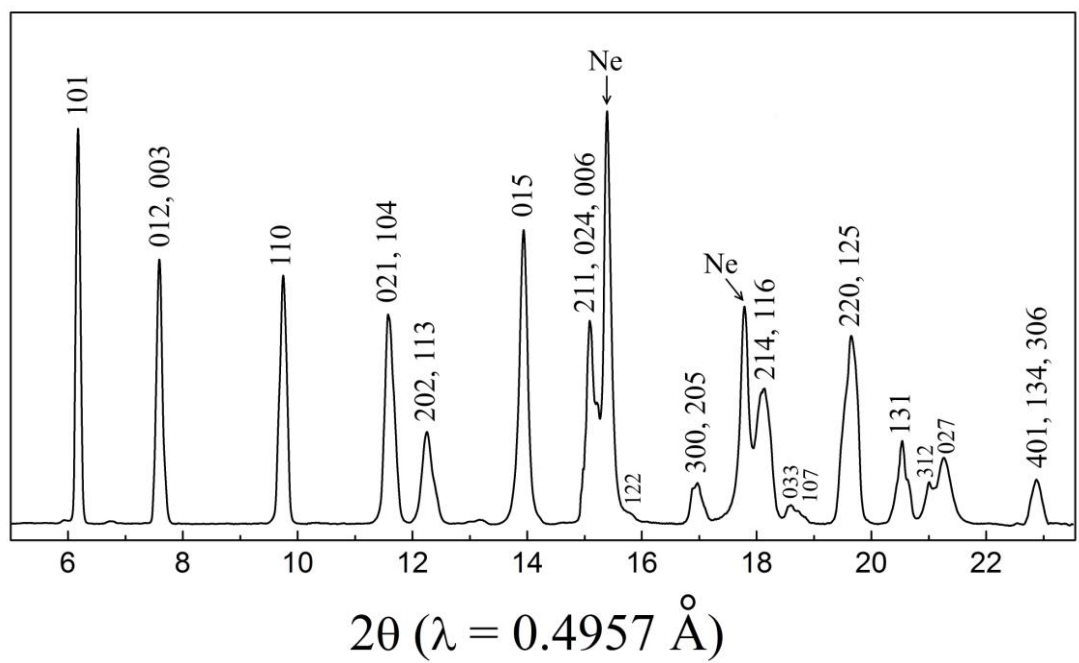

Fig. 2. Synchrotron X-ray diffraction pattern of $B_{12}As_2$ at 40.5 GPa. The *hkl* indexes of $B_{12}As_2$ lines are shown on the top.

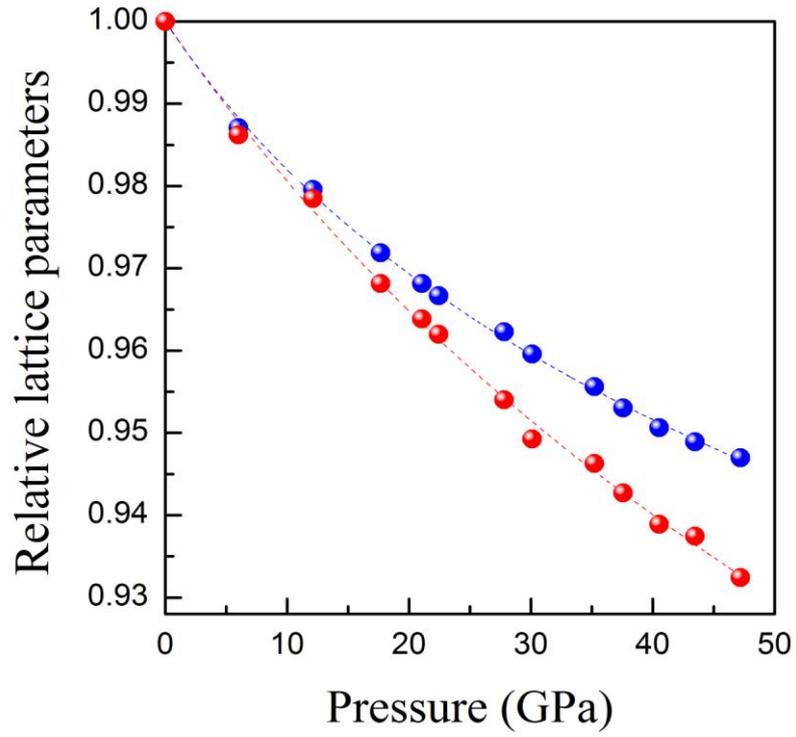

Fig. 3. Relative lattice parameters $a/a_0$ (blue) and $c/c_0$ (red) of $B_{12}As_2$ *versus* pressure. The dashed lines represent the fits of one-dimensional analog of Murnaghan equation of state to the experimental data.

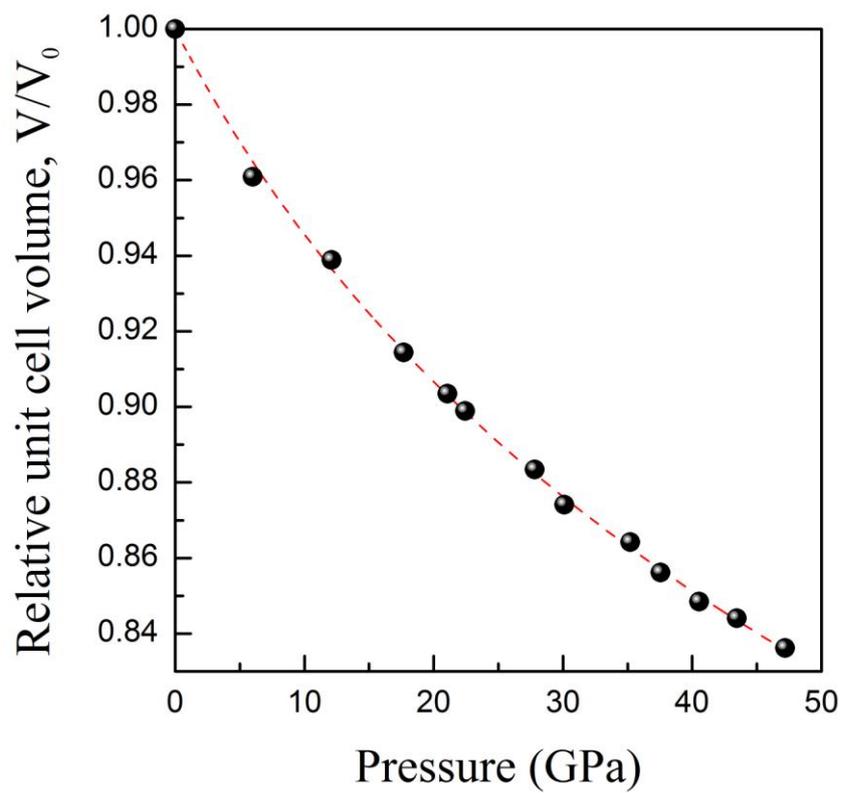

Fig. 4. Equation of state of $B_{12}As_2$. The dashed red line represents Vinet equation of state fit to the experimental data.

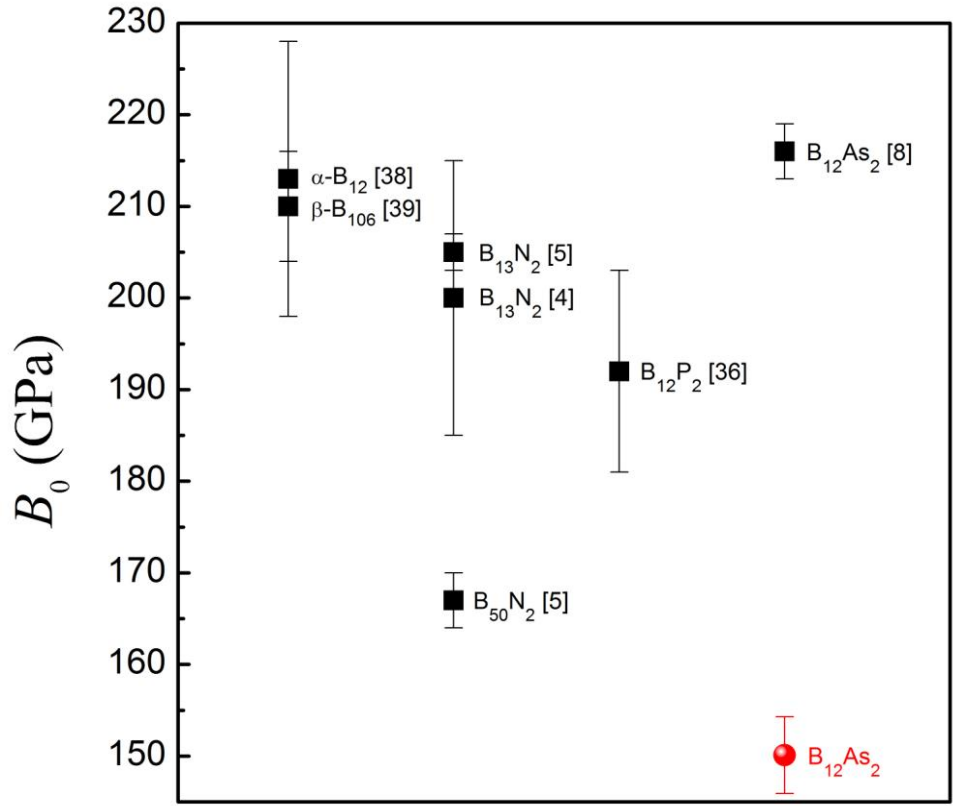

Fig. 5.  Bulk moduli of boron allotropes and boron-rich pnictides: black squares and red circle represent literature and our experimental data, respectively.